# The Shift in brain-state induced by tDCS: an EEG study

Milena Čukić[1,2*], Miodrag Stokić[3,4], Slavoljub Radenković[5], Miloš Ljubisavljević[6] and Dragoljub Donald Pokrajac[7]

[1]Department for General Physiology and Biophysics, University of Belgrade, Belgrade, Serbia

[2]Amsterdam Health and Technology Institute, HealthInc, Amsterdam, the Netherlands

[3]Life Activities Advancement Center, Belgrade, Serbia

[4]Institute for Experimental Phonetics and Speech Pathology, Belgrade, Serbia

[5] TomTom, Amsterdam, The Netherlands

[6]Department of Neurophysiology, College of Medicine and Health Sciences, UAE University, Al Ain, UAE

[7]Office of Institutional Effectiveness, Delaware State University, USA

## Abstract

Transcranial direct current stimulation (tDCS) is known to have a modulatory effect on neural tissue and that it is polarity specific. It is also shown that tDCS demonstrated the lasting effect in therapeutic applications. The main aim of the study was to examine the effects of tDCS on cortical dynamics by analyzing EEG recordings. We applied here measures taken from Recurrence Quantification Analysis, Mean State Shift (MSS) and State Variance (SV) which were previously used to detect changes in brain-state dynamics after TMS. Here, we re-used the EEG recordings from Pelliciari et al (2013) and repeated the procedure from Mutanen et al (2013). The studied cohort comprised of 16 healthy subjects; all subjects received anodal and cathodal tDCS, which were given in two separate sessions on the same day. The EEG was recorded from 10 electrodes corresponding to 10/20 standard; around left motor cortex and mirroring right cortex. The last two electrodes were P3 and P4. From three traces of recordings (pre, post1 and post2/before the stimulation, immediately after and 30min after tDCS) we extracted five different intervals (T1-T5) comprising of 500 samples. After calculating MSS and SV on those epochs and statistical testing for a significant difference, we applied Principal Component Analysis (PCA) on the same time series to check whether the data are separable. The results show that tDCS exert polarity specific effects on the MSS as shown by significantly lower MSS values after cathodal stimulation compared to anodal stimulation. Cathodal stimulation affected the SV, as compared to anodal stimulation, which did not lead to detectable changes. We are offering here for the first time an informative PCA visualization of a time-effect of a tDCS stimulation on brain state shift. Further research is needed to elucidate for how long that change can be detected and what neurobiological changes are introduced by that phenomena.

**Key words:** Transcranial direct current stimulation (tDCS), Global recurrence analysis, Principal component analysis (PCA), Electroencephalogram (EEG)

**Introduction**

Transcranial direct current stimulation (tDCS) is a well-established non-invasive brain stimulation (NIBS) technique widely used in basic neuroscience research and some clinical settings. tDCS generates a weak electrical current, ranging between 1 to 2 mA, between two electrodes, generally located on the scalp (Nitsche et al., 2008; Woods et al, 2016; Fertonani & Miniussi, 2017). tDCS modulates the neuronal resting membrane potential and spontaneous neuronal activity, in a polarity dependent manner, inducing either increase in excitability (anode) or decrease in excitability (cathode) in specific brain regions (Purpura & McMurtry, 1965; Creutzfeldt et al., 1962; Bindman et al., 1964a; Nitsche et al., 2003, 2004, 2008; Stagg and Nitsche 2011). It is reasonably well established that the effects of tDCS are associated with a number of different mechanisms, including local changes in ionic concentrations (hydrogen, calcium), alterations in protein synthesis, and modulation of N-methyl-D-aspartate (NMDA) receptor efficacy (Islam et al., 1995; Nitsche et al., 2004; Merzagora et al., 2010). Nevertheless, the effects of tDCS on a brain, on system network level are still poorly understood. Electroencephalography (EEG) provides a precise millisecond-timescale temporal dynamics to measure the postsynaptic activity in the neocortex. Furthermore, it has allowed new broad insights into the mechanism of action of tDCS and NIBS in probing and modulating neural processes, like the brain's cortical excitability, connectivity, and instantaneous state (Ilmoniemi & Kicic, 2010). The application of NIBS-EEG techniques for elucidation of changes in cortical dynamics stems from the notion that the EEG signal is a low-dimensional projection of the original primary current induced in the brain (Ilmoniemi & Kicic, 2010) by application of a stimulation/artificial disruption. If we presume that the state of the system in one particular moment is described as a point in state-space, many points from moment to moment are defining the so-called trajectory of a system (every point can be positioned with a vector starting from the beginning of the coordinate system and is representing the primary current). The lead field depending of geometry of a set-up and the conductivity of a head (Ilmoneimi & Kicic, 2010) describe how efficiently particular channel detects primary current at some coordinate (EEG is one-dimensional projection of it). If a stimulus cause the shift of the system to another part of a state-space, then any significant difference between the signal vectors (in phase-space) would also indicate a difference in original state vectors (which are describing the dynamics of the whole system) (Ilmoniemi & Kičić, 2010). That difference can be measured by various mathematical tools. Nevertheless, despite technical and methodological advance in NIBS-EEG applications the elucidation of how NIBS exactly affects underlying brain dynamics is not well characterized. Recently, Mutanen et al. (2013) introduced two recurrence quantification analysis (RQA) measures of EEG data, called mean state shift (MSS) and state variance (SV) in an attempt to probe the effects of transcranial Magnetic stimulation (TMS), another NIBS technique, on brain-state dynamics. They argue that MSS quantifies the immediate changes in the brain state after stimulation, whereas SV quantifies whether the rate at which stimulation modulates the brain state changes. They showed that the increase in MSS after stimulation implies that the brain activity occupies different regions of the brain's state space when compared to spontaneous activity. Furthermore, they suggested that these two quantitative measures could be used to study the brain dynamics affected by any stimulation irrespective of the current distribution in the brain.

In this paper, this method will be used to measure the changes of EEG complexity after tDCS. It can be argued that if SV changes after tDCS this will confirm that tDCS has the capacity to alter cortical neural network dynamics. Furthermore, its changes are polarity dependent and this may further argue that tDCS effects may be related also to some kind of at least transitory excitatory/inhibitory rebalance. Thus, we aimed to investigate the EEG changes induced by anodal and cathodal tDCS over the motor cortex during the resting state. We assume that tDCS induces a shift to a different, higher energy state which has less probability to occur in normal-spontaneous functioning mode. Furthermore, we assume that these energy-state changes may be associated with specific shifts in brain dynamics, which could be captured by examining changes of EEG complexity. If so, it would also imply that after some time the system (i.e., brain) should move back from higher to a lower energy, more probable state (according to the concept of entropy) and thus different dynamics. In the context of entropy, a high entropy value implies the ability to store more information within a neural network. Finally, to verify whether the EEG data differ in and are separable in phase-state space we complemented the analysis using the Principal Component Analysis (PCA) regularly used in our data mining projects as a feature extraction technique.

**Materials and Methods**

*Subjects*

In this study, the data obtained from the group of healthy volunteers reported earlier were analyzed focusing on changes in EEG complexity after tDCS, which was not investigated in the previous study (Pellicciari et al., 2013). The studied cohort comprised of 16 healthy subjects (8 males, eight females, 23.2 ±3 years), taken from the Pellicciari et al 2013 paper, who did not have any history of neurological diseases nor were taking CNS-active medication. All the participants were right-handed, and all gave written informed consent to participate in the study. The protocol was approved by the Ethics Committee of IRCCS Centro San Giovani di Dio Fatebenefratelli, Brescia, Italy.

*The connection between the brain state and tDCS-EEG*

From electromagnetic theory (Surutka, 1986.) current density is given as a vector function $\vec{J}(x, y, z)$ which is defined at every point in a conductive medium where an electric field exists. Its direction is the same as the current flow at the point of interest and its amplitude is given by the current divided by the area perpendicular to the flow (it is infinitesimally small). The current density is calculated from the electric field $\vec{E}$ by using the relation $\vec{J} = \gamma\vec{E}$ (Ohms law in its local form) where γ is the electric conductivity of the tissue. Additionally, the electric field is determined by the spatial gradient (summa of differential derivatives to all three directions x, y and z) of the electric potential, φ, i.e. $\vec{E} = -\nabla\varphi$. Finally, the potential inside the conductive medium is obtained by solving the continuity equation, $\nabla \cdot (\gamma\nabla\varphi) = 0$. It is a subject to the appropriate boundary conditions. Miranda et al (2009) calculated the current density distribution in two different volume conductors (cylindrical conductor model) for various electrode configurations, using finite element package to solve the continuity equation numerically, and found that the current density shows the nonlinear dependency of factors included. Rahman et al (2013) showed that tangential and radial component of the electric field has different contributions; radial (radial to the axis of a neural fiber) component seem to modulate synaptic efficacy independent of the synaptic pathway, contrary to a tangential component which seem to modulate synaptic efficacy in a pathway-specific manner. A very important segment of this process of effective stimulation is a propagation of the change induced by tDCS in the tissue. From the application of cable equation in biophysics it is well established that both of them influence Rattay's activation function (right side of original cable equation, f(x)), very important for understanding the exact influence of tDCS on the cortical tissue.

Since the brain as a very complex system, it is obeying the principle of parsimony. The state it occupies in such normal resting situation is in low (minimal) energy state which has a high probability. Then, when a stimulus is delivered, it will move to another less probable state, with increased energy. Our presumption based on Mutanen et al (2013) is that the extent of that shift in case of tDCS would be less prominent than in case of TMS which deliver the much higher amount of energy to the brain. Whatever the disturbance the system is upon, it can lead to a shift to a higher energy state which has less probability to occur in normal functioning mode. In case of the electrical field (induced by tDCS), $w_e = \frac{1}{2} \xi E^2$ where $w$ is the density of energy in tissue volume, $E$ is electric field strength, $\varepsilon$ is dielectric permeability of a medium (dielectric constant). The changes in complexity in EEG are the consequence of that state shift, as well. After some time it would be normal that the system (i.e., brain) is moving back to higher probability state with lower energy (according to the concept of entropy).

### *tDCS*

All subjects received anodal (AtDCS) and cathodal (CtDCS) tDCS, which were given in two separate sessions (morning and afternoon) on the same day. The schedule was kept constant across participants (11:30 am and 3:30 pm) to control for potential circadian effects (Sale et al., 2007). The tDCS was applied with the active (anodal or cathodal) electrode positioned over the left motor cortex corresponding to the motor representation field of the right first dorsal interosseous muscle (FDI), which was determined by TMS. The reference electrode was placed above the contralateral eyebrow (serving as a return electrode). The electrodes were 25cm$^2$ (pad size 4 x 5 cm) and were soaked in a standard saline solution (NaCl 0.9%). Electrodes were connected to a direct current stimulator (NeuroConn GmbH, Ilmenau, Germany) which delivered 1mA current, keeping the current density at 0.04mA/cm2. The duration of stimulation was 13 minutes, with a ramp-up, -down lasting 8 seconds. At the end of each stimulation session, the subjects were asked whether they felt any discomfort or other unpleasant sensations (i.e. pain). Also, subjects were asked whether they experienced any adverse effects after the previous stimulation session. No adverse effects were recorded. The researcher performing the analysis was blinded for the stimulation (i.e. conditions were coded).

### *EEG recording and data extraction*

The EEG was recorded from 10 electrodes, 4 of which were positioned around the left motor cortex corresponding to a standard 10/20 EEG electrode position nomenclature. Other four electrodes were positioned over the right hemisphere, mirroring the contralateral set-up. The last two electrodes were placed over P3 and P4. The ground electrode was placed in mid-occipital (Oz) position, whereas the right mastoid electrode served as a reference for all electrodes. The EEG signal was recorded at the sampling rate of 5 KHz, and band-pass filtered at 0.1 to 1000Hz. The skin/electrode impedance was kept below 5kΩ. All recordings were visually inspected for the presence of artifacts and only artifact-free epochs were selected for further analysis.

EEG was recorded on average for approximately 3 minutes before (pre), immediately after (post 1) and 30 minutes after tDCS (post 2). Five different time intervals (T1-5) were then selected for analysis each lasting 100 ms i.e. 500 samples. We decided to choose those intervals in order to repeat the Method Mutanen et al. (2013) used in their study. The T1 interval was sampled from the base level EEG preceding tDCS, T2 and T3 intervals from the EEG recorded immediately after tDCS (T2=1min and T3=2 min concerning the end of tDCS) and T4 and T5 intervals from the EEG recorded 30 minutes after tDCS (T4=31 min and T5=32min). Tb1 and Tb2 were used for baseline corrections (extracted from the trace before the stimulation t0/pre with the same number of samples, 500) the same as in Mutanen et al (2013) study. Thus, a data matrix comprising the raw EEG signal (i.e. voltage in μV), with two tDCS polarities (i.e. AtDCS and CtDCS) for each electrode (10) and all subjects (n=16) was created for further analysis. In this study, only EEG data recorded while subjects were instructed to keep open eyes were used for further analysis.

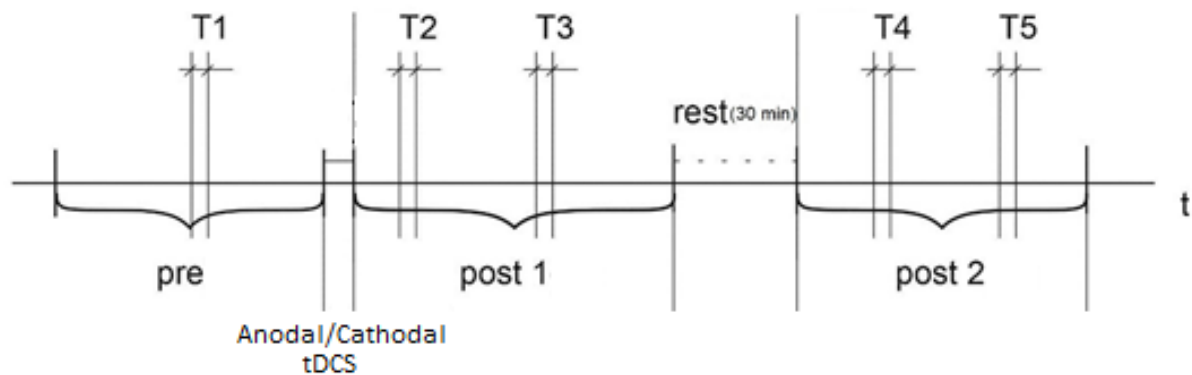

**Figure 1**: Time course of the protocol and time intervals sampled for further analysis.

## *Data analysis*

To elucidate the impact that both anodal and cathodal tDCS left on the brain during the more than a half an hour after the stimulation was over, we utilized several different methods of analysis. First we applied General Recurrence analysis and calculated Mean State Shift (MSS) and State Variance (SV) from the samples extracted from recorded EEG exactly the same as in study by Mutanen et al. (2013). Both measures are calculated from unaveraged EEG data. For all the calculations an in-house program written in Java was utilized, inspired by original code in Matlab (Mutanen et al., 2013). The only difference was that their sampling rate of recorded EEG was lower, so we were operating with five times more samples in our case (our sampling rate was 5 KHz). On those calculated measures (MSS and SV for anodal and cathodal tDCS, for all the above-mentioned epochs from T1 to T5) we performed Statistical analysis in

 IBM SPSS Statistics 17.0 for Windows (IBM Corp.) Data were summarized as means +/- standard deviation (SD). A significance level of p<0.05 (two-tailed) was used. Based on results of that part of analysis (General Recurrence Analysis) we decided to perform a simple technique we regularly use in our data mining analysis to check whether the data are separable, Principal Component analysis (PCA), but not on calculated MSS and SV values, but on original samples extracted from above defined epochs of recorded raw EEG.  PCA was used here to visualize a ten-dimensional signal (electrodes) in five observed time periods (Joliffe, 2002). PCA was performed by a script written in Matlab (version 2015b, MathWorks, Massachusetts).

*Global recurrence analysis*

Recurrence was firstly introduced by Henri Poencaré in 1890. Eckmann et al (1987) introduced a method capable of visualizing recurrence states of dynamical systems, namely recurrence plots (RP). Eckmann stated that a recurrence plot describes natural time correlation information; complex system re-visit the previous states but not in the same fashion/not identical in the strict sense. If we observe a complex system it could be described with the series of vectors $x_i$ representing its trajectory in theoretical multidimensional state space. Then the corresponding recurrence could be described by RP based on recurrence matrix (since the system is not repeating the same previous state, but just approximately). It is also important to note that equations (with which one describes such a complex system) does not explicitly comprise time; we are analysing time series, i.e. samples of EEG signal. The matrix compares the states of a system at times $T_i$ and $T_j$; that means that it is showing us when similar states of the system under study occur. Eckmann also suggested that the inverse of the longest diagonal of the recurrence matrix is proportional to the largest Lyapunov exponent of the system.

According to Eckmann's method from x(i) which is the i-th point of the orbit describing a dynamical system in d-dimensional space for i=1, …., N, we construct the recurrence matrix as an array of NxN. In that matrix, a dot is placed at (i, j) position whenever x(j) is sufficiently close to x(i) (in epsilon proximity of x(i), defined previously as a threshold). This plot of dots are representing which vectors are close enough (representing the states of a system) is a recurrence plot (Eckmann et al., 1987). The recurrence analysis was performed using Mean State Shift (MSS) and State Variance (SV) (Mutanen et al., 2013). MSS describes the mean distance between state vectors belonging to two different time intervals which use Euclidean distance. The purpose of MSS is to show whether there is a more dramatic average change in the state due to tDCS than due to the normal fluctuations in time series. MSS describes the mean distance between state vectors belonging to two different time intervals:

$$\mathrm{MSS} \equiv \mathrm{MSS}\,(T_i, T_j) = \frac{1}{N_i N_j} \sum_{ti \in Ti} \sum_{tj \in T_j} \left\| X(t_i) - X(t_j) \right\|$$

Which uses Euclidean distance; $T_i$ and $T_j$ are time intervals consisting of $N_i$ and $N_j$ discrete time points. All of the epochs we used for this analysis were 100ms long (according to our sampling rate, 500 samples). Then we calculated MSS (T1, T2), MSS (T1, T3), MSS (T1, T4) and MSS (T1, T5). Before averaging we calculated MSS (Tb1, Tb2) and then divided all the calculated MSS values which were averaged in order to elevate the differences between subjects.

SV is defined as:

$$\text{SV} \equiv \text{SV}(T_l) = \frac{1}{N_l} t \sum_{t_l \in T_l} \left\| X(t_l) - \hat{X}(T_l) \right\|^2, \text{ where } \quad \hat{X}(T_l) = \frac{1}{N_l} \sum_{t_l \in T_l} X(t_l).$$

Hence, MSS quantifies the immediate effect of tDCS on the brain state. MSS was analyzed by in-house written programs in Java programming language. Calculated MSS values were averaged to elevate the differences between subjects. To test the hypothesis that readings of EEG signals recorded before and after tDCS represent different brain states, we used multivariate analysis of variance (MANOVA) (Flury, 1997).

MSS quantifies the immediate effect of tDCS on the brain state, while SV measures the rate at which the state changes during the given time interval. The averaging that is performed in both calculations is due to suppression of the background activity which mask the effect of tDCS.

Since we did not use any thresholding, our analysis cannot be considered in the strictly sense Recurrence quantification analysis (RQA). Here, as well as in Mutanen et al work (2013) we utilized Global recurrence analysis (Marwain et al, 2007) or untresholded recurrence analysis (Iwanski and Bradley, 1998; Marwain et al, 2007). Real datasets are always finite and noisy (Rabinovich, 2006). Recurrence quantification analysis (RQA) (Webber and Zbilut, 1994) was employed to analyze non-linear dynamical characteristics of EEG data. From many studies it is confirmed that RQA can describe the non-linear nature of a short and non-stationary signal with noise (Zbilut et al, 1998, 2000). There is a consensus among researchers that global recurrence analysis is due to lack of thresholding even more robust than RQA; we measure here (as Mutanen et al, 2013) average distances between state vectors in tDCS-EEG data.

Among already mentioned, additional benefit of calculating MSS and SV is that they both can be computed from trial level data and averaged later (in order to depict induced changes due to a stimulation). Both measures are calculated from unaveraged EEG data. For all the calculations an in-house program written in Java was utilized, inspired by original code in Matlab (Mutanen et al., 2013).

### *Principal Component Analysis (PCA)*

PCA is a statistical procedure that uses an orthogonal transformation to convert a set of observations of possibly correlated variables into a set of values of linearly uncorrelated variables called principal components. This transformation is defined in such a way that the first principal component accounts for as much of variability in the data as possible, and each succeeding component in turn has the highest variance possible under the constraint that it is orthogonal the preceding components. The resulting vectors are an uncorrelated orthogonal basis set (Invented in 1901 by Karl Pearson). PCA is the simplest of the true eigenvector-based multivariate analyses. Its operation can be thought of as revealing the internal structure of the data in a way that best explains the variance in the data (Abdi and Wiliams, 2010). It is often used in exploratory data analyses, to resolve the problem of multidimensionality (in various machine learning tasks), to decorrelate the data, or to show the relatedness between populations.

If a multivariate dataset needs to be visualized PCA can provide a lower-dimensionality picture, a projection of the dataset which can be observed from the most informative viewpoint. We utilized that scenario here to obtain projections of trajectory in state-space (EEG signal space; the trajectories are measured only in discrete time points). EEG signal is the low-dimensional projection of the induced current distribution due to stimulation (Mutanen and Ilmoniemi, 2013). We hypothesized that if the trajectory (of a system/brain) changes due to stimulation that could be visualized from the EEG signal which is reflecting the primary electric activity.

Principal component analysis (PCA) defines a new orthogonal coordinate system that optimally describes variance in a single dataset (Hsu et al., 2008). PCA was used here to visualize a ten-dimensional signal (electrodes) in five observed time periods (Joliffe, 2002). PCA was performed by a script written in Matlab (version 2015b, MathWorks, Massachusets). The dimensions correspond to readings from electrodes F3, FC5, FC1, C3, P3, F4, FC6, FC2, C4 and P4.  For each participant, 10*10 covariance matrices of the observed signal and the eigenvectors and eigenvalues of the covariance matrices were initially determined. Subsequently the data were projected in the directions of the three eigenvectors corresponding to the largest eigenvalues. From every diagram pairs of states were extracted (for better visualization), namely T1vsT2, T1vsT3, T1vsT4 and T1vsT5. Those figures were constructed to illustrate the separability of observed states, representing state-shifts of the brain about tDCS stimulation and the time after tDCS.

## Results

### *Mean State Shift (MSS)*

Obtained results showed statistically significant difference between Anodal and Cathodal stimulation for averaged normalized MSS (normalized for the baseline-resting state before tDCS). For Anodal stimulation averaged values calculated for MSS were: MSS(Tb1, Tb2)= 269.36 (SD= 102.41 ), MSS(T1, T2)= 307.82 (SD= 108.79 ), MSS(T1, T3)= 295.97 (SD=102.91), MSS(T1, T4)= 313.88 (SD=86.61) and MSS(T1,T5)= 306.13 (SD= 102.69), In average 305.95 (SD= 97.23). For Cathodal stimulation averaged values calculated for MSS were: MSS(Tb1, Tb2)= 267.74 (SD= 77.41), MSS(T1, T2)= 260.15 (SD=73.85) , MSS(T1, T3)= 276.53 (SD=68.07) , MSS(T1, T4)= 259.14 (SD=79.23) and MSS(T1,T5)= 277.20 (SD=77.70); in average 268.26 (with SD=69.16). Averaged MSS, normalized with MSS (Tb1, Tb2) for Anodal stimulation was 1.173 (SD= 0.20) and for Cathodal stimulation 1.025 (SD= 0.17). No differences were obtained between different time intervals after stimulation. One-way ANOVA showed significant effect of tDCS on averaged, normalized MSS values: $F(1,31) = 4.666$, p=0.039. There is higher MSS value of the EEG signal after anodal stimulation when compared to cathodal. We found the difference between anodal and cathodal stimulation in two time intervals; T2 and T4 (0.032 and 0.02 respectively, compared to T1). Paired sample test showed significant difference in comparison A_MSS_T1_T2 vs C_MSS_T1_T2 (0.032) and in comparison A_MSS_T1_T4 vs C_MSS_T1_T4 (0.002). There are lower values of MSS after cathodal stimulation compared to anodal stimulation in T2 and T4.

### *State Variance (SV)*

Calculated values for SV for Anodal stimulation (averaged and normalized values calculated from all data) ranged from minimal 0.70 to maximal 1.68 (SD= 0.35). For Cathodal stimulation, SV values ranged from minimal 0.66 to maximal 2.519 (SD= 0.61).

Pairwise comparisons showed no statistically significant differences in SV after anodal tDCS stimulation. But we found a significant statistical difference in SV in the case of cathodal stimulation. There is a statistically higher values of SV after the cathodal stimulation: SV(T1) < SV(T5), SV(T2) < SV(T3), SV(T2) < SV(T5). Comparison of CT1 and CT5 (0.024), CT2 and CT3 (0.042) and borderline in CT2 and CT5 (0.052). Also, there is a statistically higher value of SV after cathodal stimulation in the T5 period (32 min after the stimulation) (comparison AT5 vs CT5, p=0.006). The comparison of anodal and cathodal stimulation showed the difference in the T5 interval, which is more than half an hour after the stimulation.

### *Trajectories in state-space*

For each subject and for anodal and cathodal stimulation principal components (PCs) corresponding to EEG signals from all five periods were initially displayed. There is a clear state shift associated with different time periods. Namely, clusters corresponding to pre-stimulation state (T1) are separated from the clusters immediately after tDCS (T2,T3), and 30 minutes later after the stimulus (T4, T5). The shift is present both for cathodal and anodal stimulation. MANOVA showed that for each subject the signal means corresponding to each of the observed five epochs are significantly different. There is sufficient evidence (p-value<0.0001) that the means are not equal, co-linear, co-planar nor spanning 3-dimensional space. The results suggest that the signal clusters corresponding to different epochs are significantly different. Hence, there is sufficient evidence of the existence of distinct states induced by the tDCS stimulation.

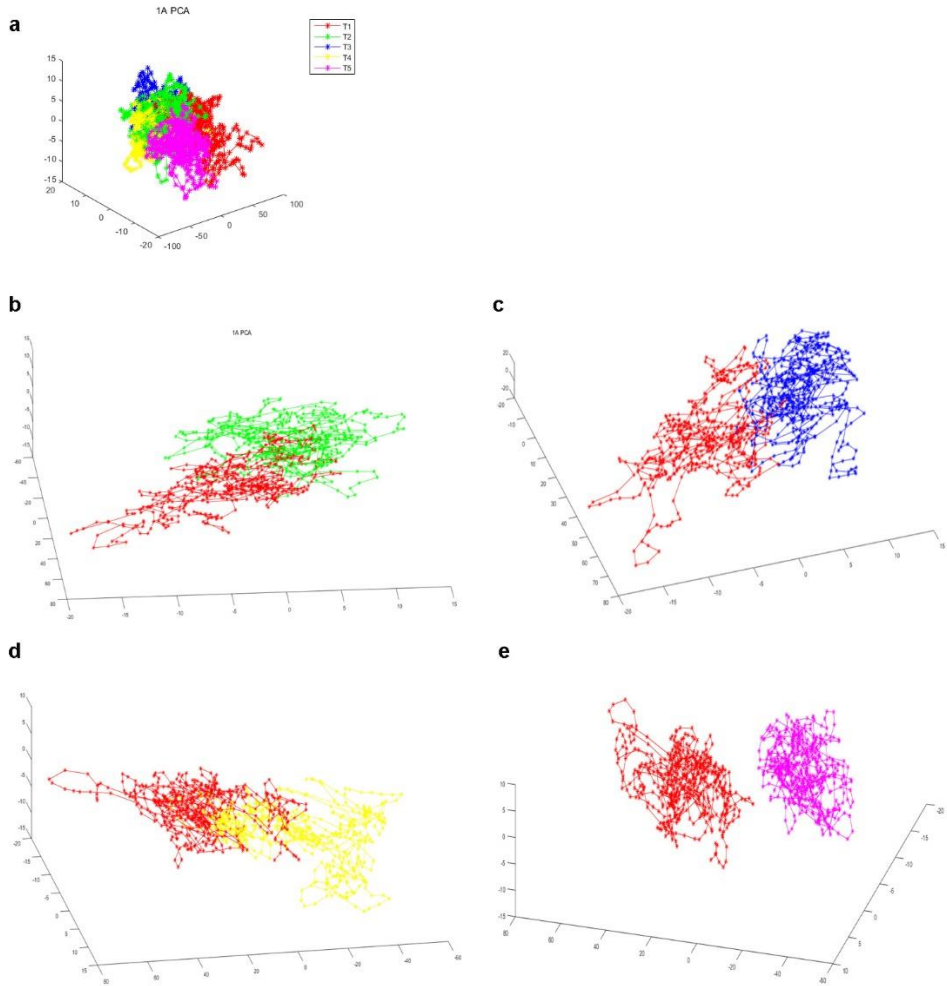

**Figure 2**: An example of the first 3 PCA components for anodal tDCS for one person's EEG; a) is composite plot of all five state trajectories, b) is depicting trajectory of T1 in red (pre stimulation) vs T2 in green (immediately after stimulation), c) T1 vs T3, d) T1 vs T4 in yellow (from t2 trace half an hour after the stimulation) and e) is showing how separated in state-space are trajectory from pre period (red) and trajectory of state T5 after 30 minutes, closer in time to the end of recording (in pink)

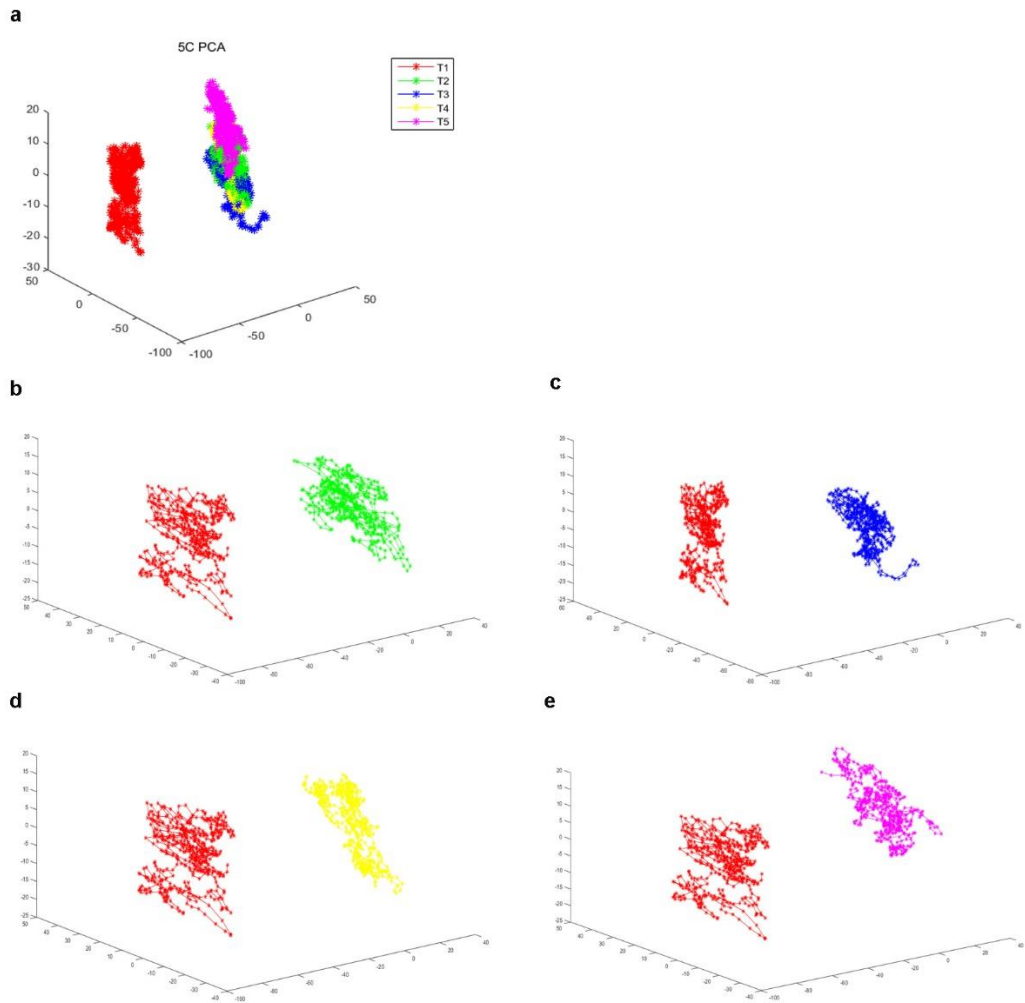

**Figure 3**. Plots of first 3 principal components (PCs) for cathodal stimulation, eyes opened condition for subject number 5. First is composite plot of all trajectories reconstructed from epochs T1, T2, T3, T4 and T5 (a). Other plots are representing separated pairs of trajectories representing different states of a brain ;T1vsT2 (b) red is trajectory from *pre* period, green is immediately *after* tDCS, T1vsT3 (c) in blue is a trajectory reconstructed from t1 interval, but not from the beginning of the trace, T1vsT4 (d) in yellow is a trajectory reconstructed from period t2, half an hour after stimulation (the beginning of the EEG trace) and T1vsT5 (e) is showing separability of states before stimulus (red) and more than half an hour after stimulus (pink); all the plots are demonstrating the change in states caused by tDCS stimulation.

In each of the five periods (T1-T5) EEG signal is represented as a sequence in ten-dimensional space. The dimensions correspond to readings from electrodes F3, FC5, FC1, C3, P3, F4, FC6, FC2, C4 and P4. We utilized MANOVA to infer the minimal dimensionality a manifold spanning the projections of means corresponding to different periods. E.g, we determine whether the means belong to a straight line, plane, etc. As a test statistic, we utilize Wilks' lambda [Flury, 1997].

Since we had those visualization for every person, for both anodal and cathodal stimulation, for all the pairs of 'states' (T1-T5), we opted to choose to represent some of them. After presenting the composite visualizations for anodal tDCS for person 1 and cathodal visualization for person 5, we are presenting here some additional figures corresponding with illustrations for persons 8, 11 and 18 to show how separable trajectories of different states look. It was impossible to present every single trajectory due to limited space here, and averaging is out of the question because every dynamic system is slightly different and we would be losing the information about the separability if we observe the group. Figures 4-6 are those representations.

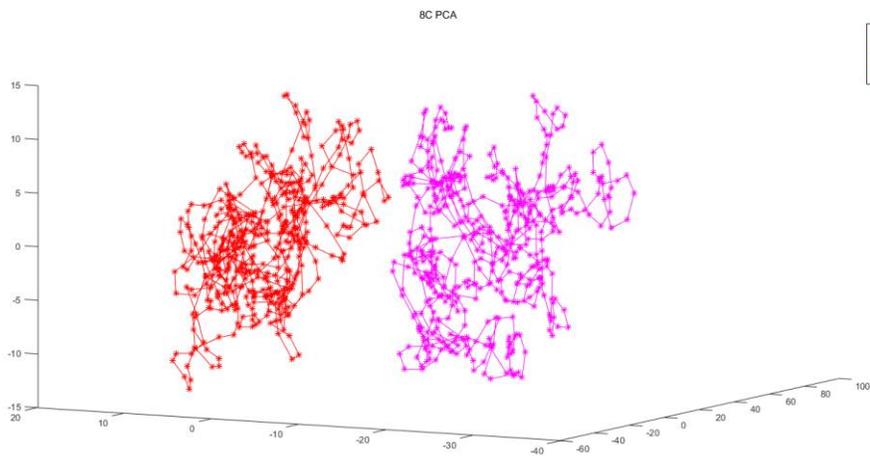

**Figure 4**: Trajectory in state-space for cathodal stimulation, person 8 (Comparison T1 and T5)

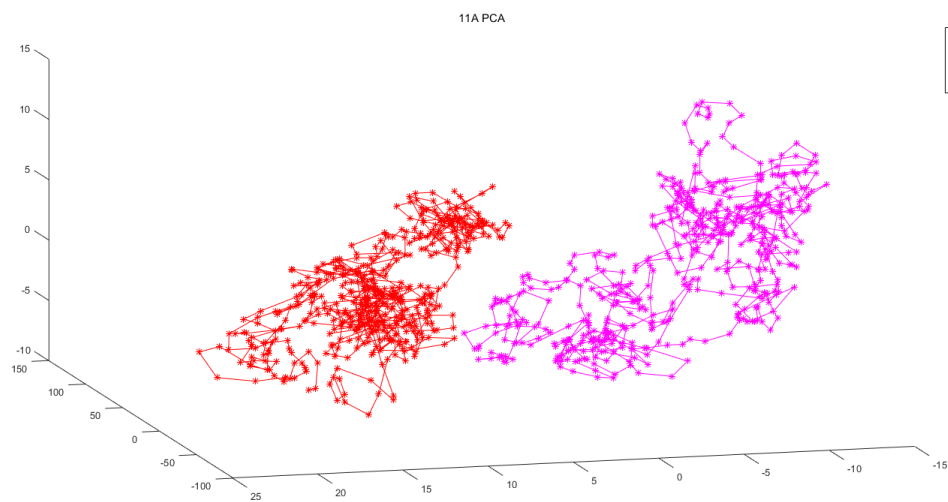

**Figure 5**: Visualization of comparison T1-T5 for anodal tDCS for person 11

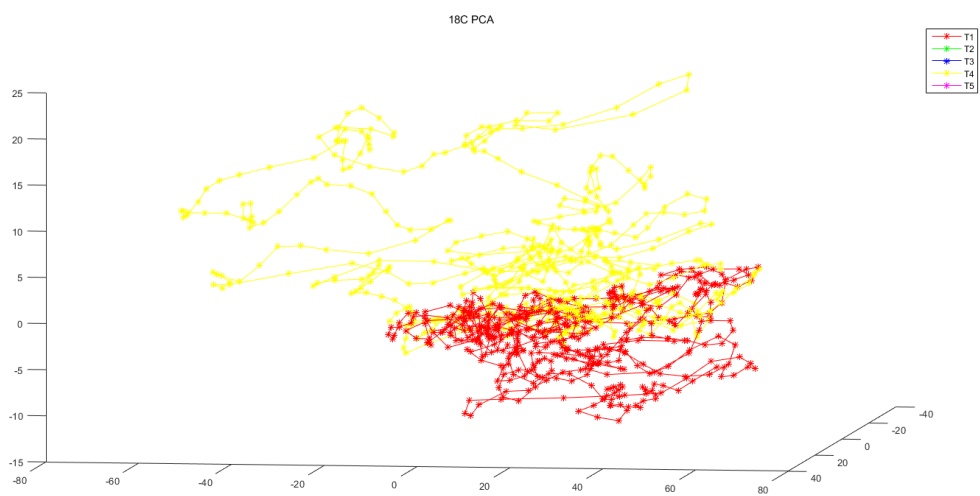

**Figure 6**: Visualization of separate trajectories of states T1 and T4, cathodal tDCS in person 18

## Discussion

The results show that tDCS exert polarity specific effects on the MSS as shown by significantly lower MSS values after cathodal stimulation compared to anodal stimulation. However, the effects were only present immediately after (T2 period) and half an hour after tDCS (T4 period). Cathodal stimulation affected the SV, as compared to anodal stimulation, which did not lead to detectable changes. Similarly, to MSS changes were present only in two intervals immediately after (T3) and 32 minutes after the stimulation (T5). Contrary to MSS and SV the PCA analysis showed the difference in brain dynamics depicted with visualization of the raw EEG samples after tDCS, which was present in all time intervals. After longer than half an hour the system is not returning to an initial state (before the onset of tDCS stimulation).

The main aim of the study was to examine the effects of tDCS on cortical dynamics. We applied MSS and SV which were previously used to detect changes in brain-state dynamics after TMS (Mutanen et al., 2013), another NIBS technique. After tDCS changes were detected in selected periods only by utilization of the same method. It should be noted that MSS quantifies the immediate changes in bran-state dynamics while SV quantifies the rate at which brain state changes. They were both significantly altered immediately after TMS. In this study the changes were present both immediately and 30 minutes after tDSC, although not in same time periods (MSS in T2 and T4 and SV in T3 and T5). Nevertheless, the results are significant as they demonstrate that tDCS can induce altered cortical dynamics outlasting the stimulation. If the system still occupies another higher less probable part of a phase-space due to the stimulation applied more than half an hour earlier, that might imply that we are looking at the proof of the capability of tDCS to induce plastic changes in the system. It is also interesting to note that the representation of that state-shift apply to all ten electrodes recordings: both to those in left motor cortex as mirroring positions on the right side cortex, as well as parietal ones. That probably imply that the effectiveness of tDCS might be considered globally seen in this experiment.

The majority of studies from the field of physiological complexity were focused of understanding the differences between healthy system and the one with a certain disorder (Eke et al, 2000; Goldberger et al, 2001; Savi, 2005; Costa el al 2006). A completely new field is forming in an attempt to understand many mental disorders by utilizing nonlinear analysis and engineer knowledge from data mining. For instance, the number of studies trying to detect nonlinear biomarkers from fMRI and EEG in depression is in such a rise that we can already call it Computational Psychiatry-when computational neuroscience is applied on psychiatry (Yahata et al, 2016; Cukic et al, 2018). We think that the results of our work can contribute to that avenue of research as well since it is already shown that physiological complexity is very important in understanding the biological underpinnings of many psychiatric disorders and we can hypothesize that our findings are getting us closer to understanding why NIBS techniques are showing to be effective in their applications in psychiatry. We might further hypothesize that from this point of view this may shed new light on understanding why tDCS showed to be efficient in treatments of medication-resistant depressions, which hallmark is elevated complexity observed in the cortex (de la Torre-Luque, 2017; Cukic et al, 2018). It should also be noted that this occurs in spite of different mechanism of tDCS action and it intensity as compared with TMS. Importantly, changes were polarity specific with cathodal stimulation causing significantly lower cortical dynamics compared to anodal stimulation. This is somewhat surprising as the majority of known behavioral effects of tDCS are associated with anodal rather than cathodal tDCS (Martin et al, 2018; Gonzalez et al, 2018). Cespon et al. (2017) found that cathodal tDCS did not show effect in improving working memory in elderly persons, and that' in elderly subjects, improved working memory after anodal tDCS applied over the left DLPFC (which)  may be related to the promotion of frontal compensatory mechanisms, which are related to attentional processes '. At present it is not clear why MVV and SV were different at different time points. One of the possible explanations may be related to the magnitude of changes in brain dynamics. From the point of view of physics and electrodynamics it is obvious that TMS and tDCS are operating on a different range of values of electrical fields and a current induced. Just to mention that according to Miranda (2006) only 50% of current in tDCS application is actually injected in the system, and the magnitude of induced field is at least two orders of value different from TMS. Laakso and his colleagues (2015) showed that inter-subject variability is in majority cases due to individual anatomical differences, where cerebrospinal fluid layer thickness variations explain 50% of individual variability. They concluded that variability in electric field induced by tDCS is related to each individual's anatomical features and can be only controlled by utilizing image processing. When the actual effect of tDCS is operating with less than micro Coulomb of imported electricity (according to Huang et al., 2017 for 2mA cortical electric field reach 0.8v/m), it is reasonable to expect that the state-shift in electromagnetic sense would be also much less pronounced than in case of TMS. We are comparing here in discussion those two modalities of NIBS just to elucidate the less pronounced effect on the brain-state shift obtained by utilization of global recurrence analysis, since our initial goal was to repeat the same methodology Mutanen et al. (2013) used on tDCS experiment. Recent study by Li et al (2018) showed that the main effect of tDCS was to acccenture default mode network (DMN) activation and salience network (SN) deactivation. Their study focused also on brain state and polarity dependent modulation of brain by tDCS by utilization of MRI. We think that further investigation is needed to compare also different montage and intensities to explore this polarity specific phenomena. Also, it would be interesting to see what is going on more precisely in other regions and in contralateral hemisphere, since we used only ten electrodes for our analysis.

To further probe into changes in dynamics we complemented the initial analysis with the PCA analysis. We are using the term 'trajectory' because we are observing here the evolution of a complex dynamical system in time, i.e. brain contrasted in states before and after the stimulation. PCA analysis is traditionally used to reduce the correlation among the data, visualization of the data and as a feature extraction in data mining techniques (Cukic et al., 2018). In this study we applied the PCA to demonstrate that it can offer additional insights into differences between brain state dynamics as it allows us to check whether the data are separable. Namely, since the loads provide the information on the separability of the data (in our case values of voltage in certain instants of time, which are measurements of EEG) PCA could help elucidate whether different groups and intervals, i.e. epochs are yielding significantly different values in reality.

Furthermore, PCA was applied on the raw data (voltages from individual EEG electrodes) and as such does not use previously calculated values nor has to do anything with recurrence plots utilized previously. Here we conducted additional statistical tests to ensure that the means are not equal, co-linear, co-planar nor spanning 3-dimensional space. This is important as it ensures that the data indeed occupy different parts of a phase-state space. We believe that this novel, previously unreported, application of PCA, allowing intuitive and critical visualization of tDCS-EEG data may provide useful marker of changes in brain dynamics related to tDCS intervention and behavioral outcome, warranting further research.

The inspection of individual PCA plots clearly showed that both anodal and cathodal stimulation induced state-shift, which was present also 30 minutes after the stimulation. It appears as if the brain dynamics shifted to a new level, which persisted for prolonged time (like, for example, the effect of stimulation in drug-resistant depression lasts for weeks). It might correspond to other researchers reporting long-time effectiveness of tDCS particularly in treatment of farmakoresistant cases of depression (Martin et al., 2018).

While we do not know the extent of time for which the system can stay in that elevated level (because of limitations of our study), that would be definitely another interesting topic to research, because it can lead us to better understanding of therapeutic effects of this stimulation.

**Conclusion**

We are offering for the first time an informative visualization of a time-effect of a tDCS stimulation on brain state shift. Although we did not succeed in repeating the RQA detection of the change between states before and after the stimulation, PCA provides us with clear separability of voltages (recording traces of EEG) induced by the action of stimulus over the course of time. Further research is needed to elucidate for how long that change can be detected and what neurobiological changes are introduced by that phenomena.

**Acknowledgements**

from the Pellicciari et al., 2013. paper) with us and for all valuable discussions. Also I want personally to thank to Tuomas Mutanen, PhD for sharing his original Matlab code for calculating MSS and SV, which we developed further and performed in Java programing language.

## References

Amassian VE, Cracco RQ, Maccabee PJ, 1989. Focal stimulation of human cerebral cortex with the magnetic coil: a comparison with electrical stimulation. Electroencephalogr. Clin Neurophysiol.74, 401-416.

Bindman LJ, Lippold OCJ, Redfearn JWT. The action of brief polarizing currents on the cerebral cortex of the rat (1) during current flow and (2) in the production of long-lasting after-effects. J Physiol (London) 1964;172:369–82.

Cespon J., Rodella C., Rossini P.M., Miniussi C., Pelliciari M.C (2017) Anodal Transcranial Direct Current Stimulation Promotes Frontal Compensatory Mechanisms in Healthy Elderly Subjects. Front. Aging Neurosci. 9:420. doi: 10.3389/fnagi.2017.00420

Costa M., Cygankiewicz I., Zareba W., Lobodzinski S. (2006). Multiscale Complexity Analysis of Heart Rate Dynamics in Heart Failure: Preliminary Findings from the MUSIC Study. Computers in cardiology 33:101 – 103.

Creutzfeldt, O. D., Fromm, G. H., Kapp, H. (1962). Influence of transcranial d-c currents on cortical neuronal activity. Exp Neurology, 5, 436-52.

Čukić M, Platiša M, Kalauzi A, Oommen J,. Ljubisavljević M. (2017). The comparison of Higuchi's fractal dimension and Sample Entropy analysis of sEMG: effects of muscle contraction intensity and TMS. Fractal Geometry and Nonlinear Analysis in Medicine and Biology (Oatext) 3(2), 2017.

Cukic M., Pokrajac D., Stokic M., Simic S, Radivojevic V, Ljubisavljevic M. (2018). EEG machine learning with Higuchi fractal dimension and Sample Entropy as features for successful detection of depression. Arxive.org/Preprint at Cornell repository for Statistics/Machine learning https://arxiv.org/abs/1803.05985.

De la Torre-Luque Alejandro and Bornas Xavier (2017) Complexity and Irregularity in the Brain Oscillations of depressive Patients: A Systematic Review. Neuropsychiatry (London)(2017)&(5), 466-477.

Eke A, Herman P, Kocsis L, Kozak LR. (2002). Fractal characterization of complexity in temporal physiological signals. Physiol Meas. 23, R1-38.

Eckmann J.P., Kamphorst S.O., and Ruelle D (1987). Recurrence plots of dynamical systems. Europhys. Lett. 4, 973-977.

Fertonani, A., and Miniussi, C. (2017). Transcranial electrical stimulation: what we know and do not know about mechanisms. Neuroscientist 23, 109–123. doi: 10.1177/1073858416631966

Flury B., A First Course in Multivariate Statistics, Springer-Verlag New York, Inc., 1997.

Friston. K. (2010) The free-energy principle: a unified brain theory? Nat. Rev. Neurosci. 11, 127-138.

Goldberger AL, Peng CK, Lipsitz LA. (2002). What is physiologic complexity and how does it change with aging and disease? Neurobiology of Aging, 23, 23-26.

Gonzalez P.C, Kenneth N. K. Fong, and Ted Brown (2018) The Effects of Transcranial Direct Current Stimulation on the Cognitive Functions in Older Adults with Mild Cognitive Impairment: A Pilot Study. Behavioural Neurology. Volume 2018, Article ID 5971385, 14 pages

https://doi.org/10.1155/2018/5971385

Huang Y, Liu A.A., Lafon B., Friedman D., Dayan M., Wang X., Bikson M., Doyle W.K., Devinsky O, Parra L.C. (2017) Measurements and models of electroc fields in the in vivo juman brain during transcranial electric stimulation. eLife 2017;6:e18834.DOI: 10.7554/eLife.18834

Ilmoniemi, R., Kičić, D. (2010). Methodology for combined TMS and EEG. Brain Topogr. 22, 233-248.

Islam,N.,Aftabuddin,M.,Moriwaki,A.,Hattori,Y.,andHori,Y.(1995a).Increase in the calcium level following anodal polarization in therat brain. BrainRes. 684,206–208.

Islam,N.,Moriwaki,A.,Hattori,Y.,Hayashi,Y.,Lu,Y.F.,andHori,Y.(1995b).c-Fos expression mediated by N-methyl-D-aspartate receptors following anodal polarization in the rat brain. Exp.Neurol. 133,25–31.doi:10.1006/exnr.1995. 1004

Iwanski J.C. and Bradley (1998) Recurrence plots of experimental data: To embed or not to embed? Chaos. 1998 Dec;8(4):861-871.

Laakso Ikka, Tanaka Satoshi, Koyama Soichiro, De Santis Valerio, Hirata Akimasa (2015) Inter-subject Variability in Electric Fields of Motor Cortical tDCS.Brain Stimulation 8 (2015) 906-913.

Li L.M., Violante I.R., Leech R., Ross E., Hampshire A., Opitz A., Rothwell J.C., Carmichael D.W, Sharp D.J. (2018) Brain state and polarity dependent modulkation of brain networks by transcranial current stimulation. Hum Brain Mapp 2018;1-12. DOI: 10.1002/hbm.24420

Mangia Anna L., Marco Pirini and Angelo Cappello (2014) Transcranial direct current stimulation and power spectral parameters: a tDCS/EEG co-registration study. Frontiers in Human neuroscience, 07 August 2014 |Volume 8 | Articel 601

Martin DM, Moffa A, Nikolin S, Bennabi D, Brunoni AR, Flannery W, Haffen E, McClintock SM, Moreno ML, Padberg F, Palm U, Loo CK.(2018) Cognitive effects of transcranial direct current stimulation treatment in patients with major depressive disorder: An individual patient data meta-analysis of randomised, sham-controlled trials. Neurosci Biobehav Rev. 2018 Jul;90:137-145. doi: 10.1016/j.neubiorev.2018.04.008. Epub 2018 Apr 13.

Marwin, N., Carmen Romano, M.,Thiel, M., and Kurths, J. (2007) Reccurence plots for the analysis of complex systems. Phys. Rep. 438, 237-329.

Miranda, P. C., Lomarev, M., Hallet, M. (2006). Modeling the current distribution during transcranial direct current stimulation. Clinical Neurophysiology, doi:10.1016/j.clinph.2006.04.009

Merzagora AC1, Foffani G, Panyavin I, Mordillo-Mateos L, Aguilar J, Onaral B, Oliviero (2010) Prefrontal hemodynamic changes produced by anodal direct current stimulation. Neuroimage. 2010 Feb 1;49(3):2304-10. doi: 10.1016/j.neuroimage.2009.10.044. Epub 2009 Oct 21.

Mutanen, T., Nieminen, J. O., Ilmoniemi, R. (2013). TMS-evoked changes in brain-state dynamics quantified by using EEG data. Frontiers in Human Neuroscience, 25 april, 2013| Volume7, Article 155, doi: 10.3389/fnhum.2013.00155.

Nitsche, M. A., Fricke, K., Henschke, U., Schlitterlau, A., Liebetanz, D., Land, N., Hening, S., Terau, F., Paulus, W. (2003). Pharmacological modulation of cortical excitability shifts induced by transcranial direct current stimulation in humans. J Neurophysiol., 553, 293-301.

Nitsche, M. A., Grundey, J., Lievetanz, D., Lang, N., Tergau, F., Paulus, W. (2004). Cateholaminergic consolidation of motor cortical neuroplasticity in humans. Cereb Cortex, 14, 1240-5.

Nitsche MA, Cohen LG, Wassermann EM, Priori A, Lang N, Antal A, Paulus W, Hummel F, Boggio PS, Fregni F, Pascual-Leone A. (2008) Transcranial direct current stimulation: State of the art 2008. Brain Stimul. 2008 Jul;1(3):206-23. doi: 10.1016/j.brs.2008.06.004. Epub 2008 Jul 1.

Ouyang, G. Li, X., Dang, C., and Richards, D.A. (2008) Using recurrence plot for determinism analysis of EEG recordings in genetic absence epilepsy rats. Clin Neurophysiol. 119, 1747-1755.

Opitz,A.,Paulus,W.,Will,S.,Antunes,A.,andThielscher,A.(2015). Determinants of the electric field during transcranial direct current stimulation. Neuroimage 109,140–150.doi:10.1016/j.neuroimage.2015.01.033

Pelliciari, M. C., Brignani, D., Miniussi, C. (2013). Excitability modulation of the motor system induced by transcranial direct current stimulation: a multimodal approach. Neuroimage, 83, 569-580.

Purpura, D. P., McMurtry, J. G. (1965). Intracellular activities and evoked potential changes during polarization of motor cortex. J. Neurophysiology, 28, 166-185.

Rahman Asim, Davide Reato, Mattia Arlotti, Fernando Gasca, Abhishek Datta. Lucas C. Parra (2013) Cellular effects of acute direct current stimulation: somatic and synaptic terminal effects. J. Physiol 591.10 (2013) pp 2563-2578

Sale M.V, Ridding M.C, Nordstrom M.A. (2007) Factors influencing the magnitude and reproducibility of corticomotor excitability changes induced by paired associative stimulation. Experimental Brain Research August 2007, Volume 181, Issue 4, pp 615–626

Savi M. (2005). Chaos and Order in Biomedical Rhythms. J Braz. Soc. Mech. Sci. & Eng. April-June 2005, Vol. XXVII, No. 2 / 157

Stagg, C. J., Nitsche, M. A. (2011). Physiological Basis of Transcranial Direct Current Stimulation. The Neuroscientist, 17(1) 37-53.

Surutka J. Osnovi elektrotehnike.Naučna knjiga Beograd. 1986.

Woods AJ, Antal A, Bikson M, Boggio PS, Brunoni AR, Celnik P, Cohen LG, Fregni F, Herrmann CS, Kappenman ES, Knotkova H, Liebetanz D, Miniussi C, Miranda PC, Paulus W, Priori A, Reato D, Stagg C, Wenderoth N, Nitsche MA (2016). A technical guide to tDCS, and related non-invasive brain stimulation tools. Clin Neurophysiol. 2016 Feb;127(2):1031-1048. doi: 10.1016/j.clinph.2015.11.012. Epub 2015 Nov 22

Yahata N., Kasai K., Kawato M. (2016). Computational neuroscience approach to biomarkers and treatments for mental disorders. PCN, https://doi.org/10.1111/pcn.12502

Zbilut J.P,Giuliani A. and.Webber C.L.Jr.(1994) Detecting deterministic signals in exceptionally noisy environments using cross-recurrence quantification. Physics Letters A, Volume 246, Issues 1–2, 7 September 1998, Pages 122-128

Zbilut J. and Webber C.L. (1985) Dynamical assessment of physiological systems and states using recurrence plot strategies. J Appl Physiol 76:965–973, 1994